\documentclass[preprint,showpacs,showkeys,preprintnumbers,amsmath,amssymb]{revtex4}

\usepackage{graphicx}
\usepackage{dcolumn}
\usepackage{bm}
\usepackage{verbatim}

\begin{document}
\newcommand{\ud}{\mathrm{d}}

\title{A pepper-pot emittance meter for low-energy heavy-ion beams}

\author{H.R.~Kremers\footnote{Author to whom correspondence should be addressed. Electronic mail: h.r.kremers@rug.nl.}, J.P.M.~Beijers, and S.~Brandenburg}

\affiliation{Kernfysisch Versneller Instituut, University of Groningen, Zernikelaan 25, 9747 AA Groningen, The Netherlands}

\date{\today}

\begin{abstract}
A novel emittance meter has been developed to measure the four-dimensional, transverse phase-space distribution of a low-energy ion beam using the pepper-pot technique. A characteristic feature of this instrument is that the pepper-pot plate, which has a linear array of holes in the vertical direction, is scanned horizontally through the ion beam. This has the advantage that the emittance can also be measured at locations along the beam line where the beam has a large horizontal divergence. A set of multi-channel plates, scintillation screen and ccd camera is used as a position-sensitive ion detector allowing a large range of beam intensities that can be handled. This paper describes the design, construction and operation of the instrument as well as the data analysis used to reconstruct the 4D phase-space distribution of an ion beam. Measurements on a 15~keV He$^+$ beam are used as an example.
\end{abstract}

\pacs{07.05.Pj, 07.77.Ka, 29.27.Fh, 41.75.Ak, 41.85.Ew, 41.85.Ja}
\keywords{Beam emittance, Pepper-pot, ECR ion source}
\maketitle


\section{\label{sec:intro}Introduction}
The continuous demand for higher beam intensities at heavy-ion accelerator facilities is particularly challenging for the injector chain consisting of one or more heavy-ion sources and a low-energy beam transport line\cite{fukumishi,machicoane,zhao}. The reason for this is that the intense, low-energy heavy-ion beams extracted from the source(s) generally have large emittances, which makes it difficult to transport and inject these beams into the accelerator with minimum beam losses. In order to minimize beam losses it is important to have a good knowledge of the full four-dimensional (4D) phase-space distribution of the ion beam including all its correlations and to keep emittance growth caused by space-charge effects, ion-optical aberrations and magnet misalignments under control. In this paper we describe a versatile emittance scanner based on the pepper-pot principle that we have developed and constructed to measure the full 4D phase-space distributions of low-energy, heavy-ion beams. We have used this device to study beam extraction from an Electron Cyclotron Resonance Ion Source (ECRIS) and its subsequent transport and injection into the AGOR cyclotron\cite{saminathan}.

Pepper-pot emittance meters are devices to measure the full 4D transversal phase-space distribution of charged-particle beams. This is an important advantage compared to alternative emittance measurement devices such as Allison scanners and the three-screen and quadrupole methods, which give much less detailed information of the phase-space distribution and only in a single transverse plane. Ion-optical aberrations can induce large correlations between the horizontal and vertical planes and these can only be measured with pepper-pot emittance meters.

The basic principle of a pepper-pot emittance meter is as follows: The ion beam to be probed is intercepted by a pepper-pot plate which has a pattern of small holes. The beamlets transmitted by these holes are projected onto a position-sensitive detector located at a fixed distance downstream of the pepper-pot plate. The hole pattern in the pepper-pot plate defines the spatial sampling of the beam's phase space, while the angular information is extracted from the images of the beamlets on the position-sensitive detector. The spatial resolution of a pepper-pot emittance meter is determined by the requirement that different beamlets do not overlap in the imaging plane, while the angular resolution is determined by the spatial resolution of the position-sensitive detector. In the first pepper-pot emittance meters kapton foils or photographic films were used as position-sensitive detectors\cite{collins,guharay}, but nowadays a scintillation screen combined with a ccd camera is used allowing direct electronic readout of the beamlet images\cite{pfister,kondrashev}. However, practical experience with the use of scintillation screens for quantitative measurements on low-energy ion beams shows that often problems arise because of inhomogeneity, nonlinear response and aging of the scintillator material.

We have chosen for a multi-channel plate (MCP), phosphor screen and ccd camera combination as a position-sensitive ion detector. This has the advantage that, because of the large dynamic range of the ion detector, our pepper-pot emittance meter can be used with very small beam currents, e.g.\ with secondary radioactive ion beams. Another difference is that our emittance meter has a pepper-pot plate with a one-dimensional array of holes instead of a 2D pattern. The 4D phase-space distribution of an ion beam is determined by stepping the pepper-pot plate through the beam in the direction perpendicular to that of the array of holes and measuring the images of the transmitted beamlets at each step. This scheme offers greater flexibility with respect to the location along a beam line where the emittance can be measured. Conventional pepper-pot devices using a 2D hole pattern can only measure beam emittances at locations along the beam line where the beam divergences are sufficiently small so that beamlets from neighbouring holes do not overlap. Our device, on the other hand, can also determine the beam emittance with good resolution at locations where the beam has a large divergence and small cross section in one transverse plane and a small divergence and large cross section in the other transverse plane. The advantage of our scheme is that the spatial resolution in the scanning direction is greatly increased and basically determined by the step size, the disadvantage that we do not measure the phase-space distribution in a single shot.

In the following we will first describe the design and construction of our pepper-pot emittance meter in section~\ref{design}. The measurement procedure and phase-space reconstruction are described in section~\ref{procedure} and the paper ends with some concluding remarks in
section~\ref{conclusion}.


\section{\label{design}Design aspects}
\begin{figure}[tb!]
\centering
\includegraphics[width=7.5cm]{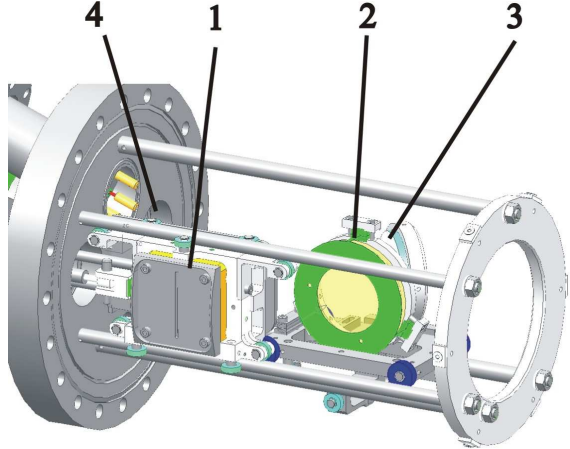}
\caption{\label{fig:pepperpot}The pepper-pot emittance meter with its cylindrical shield removed. 1) Pepper-pot plate, 2) MCP-phosphor screen
combination, 3) flat mirror and 4) window. The lens and ccd camera (not shown) are located outside the vacuum.}
\end{figure}

\begin{figure}[tb!]
\centering
\includegraphics[width=10.5cm]{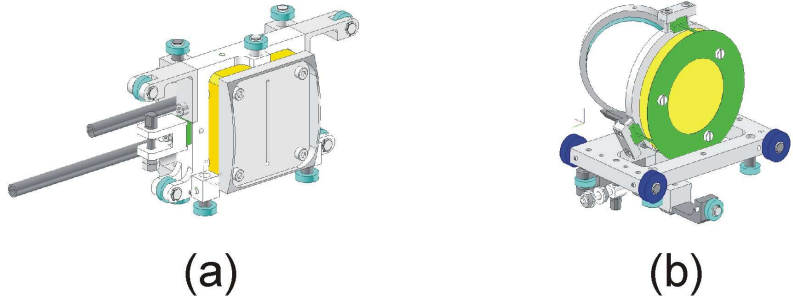}
\caption{\label{fig:tables}(a) Translation table with pepper-pot plate. (b) Translation table with MCP-phosphor screen-mirror assembly.}
\end{figure}

An overall view of the pepper-pot emittance meter is shown in Fig.~\ref{fig:pepperpot}. The pepper-pot plate and ion detector assembly are both attached to translation stages and shown in more detail in Fig.~\ref{fig:tables}. The table with the pepper-pot plate can be moved horizontally through the beam with a maximum range of 110~mm and a position accuracy better than 10~$\mu$m using a stepper motor mounted outside the vacuum. The pepper-pot plate is a tantalum foil with a thickness of 25~$\mu$m clamped on a water-cooled copper block which can absorb a maximum beam power of 150~W. The plate has a vertical row of 20 holes with diameters of 20~$\mu$m each and a pitch of 2~mm. The hole areas have been optically calibrated in order to normalize the images of the transmitted beamlets in the phase-space reconstruction algorithm (see section~\ref{procedure}).

The position-sensitive ion detector consists of two MCP's in chevron configuration combined with a phosphor screen mounted 52~mm behind the pepper-pot plate. The total gain of the MCP's is determined by the bias voltage which is set, depending on the ion current, at such a value as to not saturate the ccd camera. The maximum gain is $1.8\cdot 10^8$ and warrants a very large dynamic range for the beam intensity, i.e.\ from single ions up to approximately $3\cdot 10^{15}$~s$^{-1}$ (500~$\mu$A). The electron clouds emerging from the MCP's are accelerated towards the phosphor screen by a strong electric field. The effective diameter of the MCP-phosphor screen combination is 41.5~mm. The phosphor screen emits photons with a wavelength of approximately 550~nm and intensity proportional to the incident ion flux on the MCP's. The photons are reflected by an aluminum mirror with a diameter of 50~mm mounted under 45~degrees behind the phosphor screen, exit the vacuum chamber through a 3~mm thick glass window and are focused by a lens system with magnification factor $M=0.1$ onto the ccd camera. The MCP's, phosphor screen and mirror assembly are mounted on a translation table which can be moved in and out of the beam pneumatically. A cylindrical metal shield with a rectangular opening for the beam protects the ion detector from stray and/or reflected charged particles. The ccd camera measures the light-intensity distribution on $1236\times 1628$ square pixels with a size of $4.4\,\mu$m. Ccd images are electronically stored in $1236\times 1628$ integer arrays $g_{ij}$ of grey values in the range between 0 and 1023. The exposure time can be set between $10^{-4}$ and $10^{-1}$~s, normally a value of 1/30~s is used. The time resolution is limited by the decay time of the phosphor fluorescence, which is $\approx 1.2$~ms in our case. If needed, one can use a faster phosphor to improve the time resolution. The entire emittance meter is mounted on a 150~mm diameter conflat flange and extends 255~mm inside the vacuum chamber. Finally, all components of the device that are mounted inside the vacuum can be baked at a maximum temperature of 300~degree Celsius.

\begin{figure}[tb!]
\centering
\includegraphics[width=12.cm]{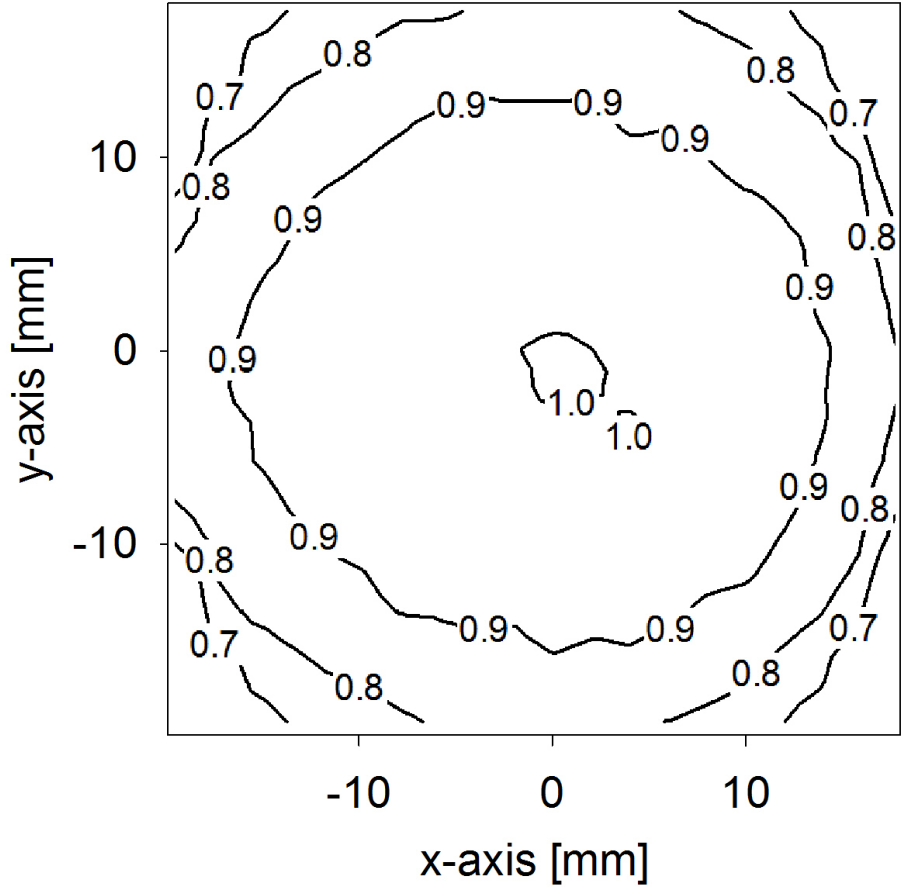}
\caption{\label{fig:screm}Intensity contours of a ccd image of a uniformly-illuminated light source at the position of the phosphor screen. The contours show a slight left-right asymmetry caused by the 45~degree mirror.}
\end{figure}

In order to check the quality of the pepper-pot plate and imaging system we have performed two test measurements. In the first one the pepper-pot plate was mounted at the position of the phosphor screen and imaged onto the ccd camera using a parallel and uniform light beam. The equidistant holes in the pepper-pot plate were imaged onto single and equidistant pixels in the same column of the ccd chip. From this image the spatial calibration of the ccd camera has been derived: 46~pixels correspond to 2~mm so that one pixel length or width corresponds to $2000/46=43.48\;\mu$m in the pepper-pot plane. In the second test measurement we mounted a 2D pepper-pot mask with a square $20\times 20$ array of 1~mm diameter holes with a pitch of 2~mm at the position of the phosphor screen and illuminated the mask uniformly. A contour plot of the smoothed and normalized ccd image, back transformed to the plane of the phosphor screen, is shown in Fig.~\ref{fig:screm}. It represents a 2D correction function for the intensity decrease of the ccd image towards its edges due to the finite solid angle of the optical system. The correction function is not cylindrically symmetric around the optical axis. This asymmetry is caused by the 45 degree mirror and is taken into account in the reconstruction of phase-space from the pepper-pot images.


\section{\label{procedure}Measurement procedure and phase-space reconstruction}
A LabVIEW\texttrademark program has been written to control the pepper-pot emittance meter, set relevant measurement parameters, perform an emittance measurement and reconstruct the 4D phase-space distribution of the ion beam\cite{labview}. In this section we describe the emittance measurement procedure, the algorithm used to reconstruct the phase-space distribution $\rho(x,y,x',y')$ from the pepper-pot data and the calculation of beam emittances.

\subsection{Emittance measurement procedure}
Each emittance measurement starts with a calibration of the position of the pepper-pot plate by moving the plate carefully against a calibration switch. The pepper-pot plate and ion detector/mirror are then positioned in the center of the beam line and exposed to the ion beam to set the voltage on the MCP's. This is done by taking ccd images of the transmitted beamlets and setting the voltage at such a value that the most illuminated pixels stay well below saturation. Next a test scan is performed to determine the width of the beam and to set the horizontal scan range. The test scan is also used to determine the number of pixel rows and columns that will be stored at each horizontal step during an emittance measurement. Since the transmitted beamlets all lie in a narrow vertical strip of which the width depends on the horizontal beam divergence, we do not store the entire $1236\times 1628$ grey-value array $g_{ij}$ at each step. Instead a $i_{max}\times j_{max}$ sub-array is stored with the central column shifting synchronously with the position of the pepper-pot plate. Typical values for $i_{max}$ and $j_{max}$ are $i_{max}=920$ and $j_{max}=240$. The last step before an emittance scan can be performed is to measure a so-called empty-frame image for background subtraction. This is done by averaging 30 ccd images with the ion beam being blocked upstream. The resulting background image $bg_{ij}$ has a grey-value distribution determined by dark-current and fixed-pattern noise. Dark-current noise is Gaussian, while the fixed-pattern noise is non-Gaussian showing the microscopic structure of the ccd chip. After these preparatory steps the pepper-pot plate is moved to the start position and an emittance measurement can be started.

\begin{figure}[tb!]
\centering
\includegraphics[width=14.cm]{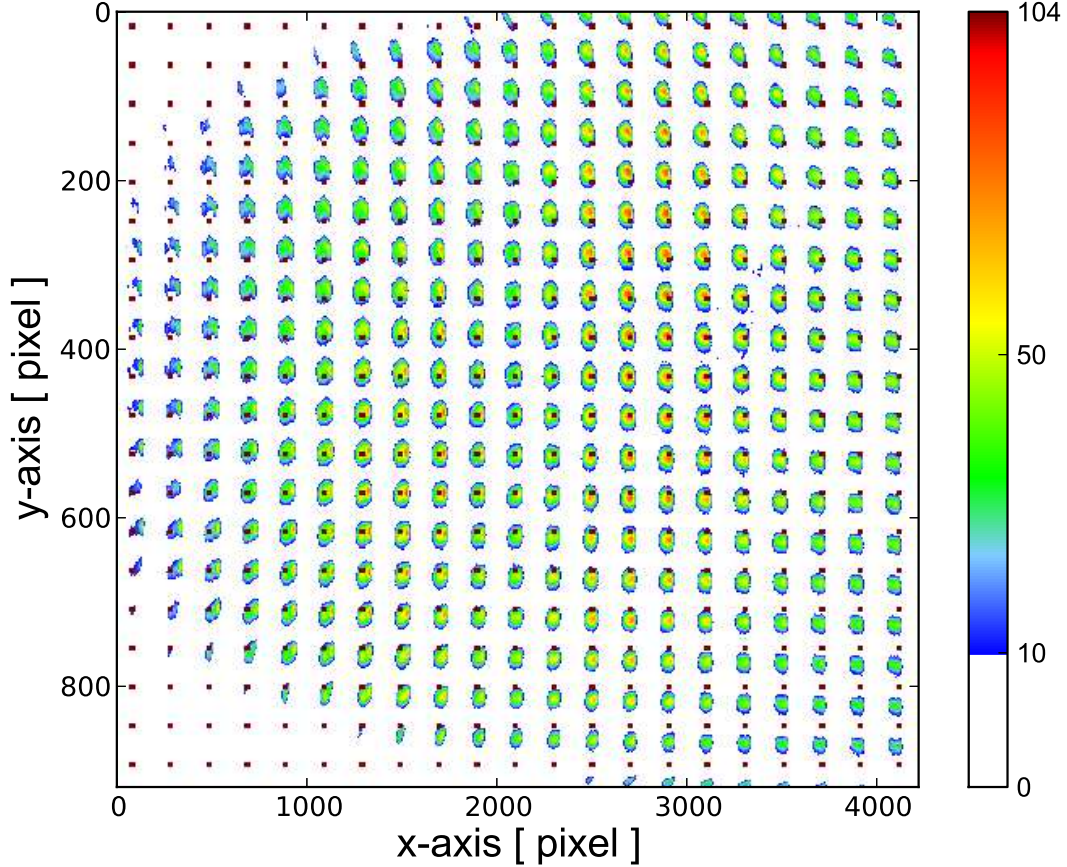}
\caption{\label{fig:concat}Concatenated pepper-pot data image of a measurement on a 15~keV He$^+$ beam with $N=21$ horizontal steps and step size of 1~mm from which the 4D phase-space distribution is reconstructed. The black dots indicate the position of the holes in the pepper-pot plate.}
\end{figure}

An emittance measurement is performed by scanning the pepper-pot plate in $N$ steps through the ion beam and taking a ccd image of the transmitted beamlets at each step. After subtracting the background array $bg_{ij}$ from each image the $N$ $i_{max}\times j_{max}$ sub-arrays are concatenated into a single data array $G_{ij}$, with the array indices $i$ and $j$ running from 1 to $i_{max}$ and from 1 to $Nj_{max}$, respectively. The $i_{max}\times Nj_{max}$ array $G_{ij}$ contains all the data from which the 4D transversal phase-space distribution $\rho(x,y,x',y')$ of the ion beam can be reconstructed. An example of such a data array $G_{ij}$ obtained from a measurement on a 15~keV He$^+$ beam performed at the EIS test bench of GSI \cite{eis-gsi} is shown in Fig.~\ref{fig:concat}.

\subsection{Phase-space reconstruction}
To reconstruct the phase-space distribution $\rho(x,y,x',y')$ we have to extract from the pepper-pot data file $G_{ij}$ the phase-space density $\rho$ at the phase-space coordinates $(x,y,x',y')$. This can only be done unambiguously if beamlets transmitted by neighbouring holes in the pepper-pot plate do not overlap in the detector plane. When this is the case we can attach to all array elements $G_{ij}$ corresponding to a particular beamlet spot upper-indices $(p,q)$ which represent the pixel coordinates of the pepper-pot hole from which the beamlet emerged, i.e.\ $G_{ij}^{pq}$. The pixel coordinates $(p,q)$ determine the spatial coordinates $(x_q,y_p)$ in the pepper-pot plane at which the phase-space distribution $\rho(x,y,x',y')$ is sampled:
\begin{equation}
x_q=\frac{(q\bmod{j_{max}})\Delta xp}{M}+\lfloor q/j_{max}\rfloor\Delta xs \hspace{0.8cm}; \hspace{0.8cm}y_p=\frac{p\Delta yp}{M}\;, \label{eq:detplancor}
\end{equation}
with $\Delta xp=\Delta yp=4.4\cdot 10^{-3}$~mm the linear size of the ccd pixels, $M=0.1$ the linear magnification of the optical system which images the phosphor screen onto the ccd camera, $\lfloor\cdot\rfloor$ the floor function and $\Delta xs$ the step size with which the pepper-pot plate is scanned through the ion beam. The direction tangents $(x'_{jq},y'_{ip})$ at each spatial position $(x_q,y_p)$ in the pepper-pot plane is determined by the pixel coordinates $(i,j)$ in the ccd image of the beamlet that emerged from the hole in the pepper-pot plate with pixel coordinates $(p,q)$ via
\begin{equation}
x'_{jq}=\frac{(j-q)\Delta xp}{Md}\hspace{0.8cm};\hspace{0.8cm}y'_{ip}=\frac{(i-p)\Delta yp}{Md}\;, \label{eq:detplanang}
\end{equation}
with $d=52$~mm the drift length between the pepper-pot plate and the ion detector.

Equations~(\ref{eq:detplancor}) and (\ref{eq:detplanang}) relate the set of discrete coordinates $(x_q,y_p,x'_{jq},y'_{ip})$ at which the phase-space density $\rho$ is sampled to the lower- and upper-indices of the grey-value array $G_{ij}^{pq}$. They also determine the spatial and angular resolutions that can be obtained. The horizontal spatial resolution $\Delta x$ is either given by $\Delta xp/M=44\cdot 10^{-3}$~mm or by the horizontal step size $\Delta xs$, whichever is largest. In practice it is the step size that determines the horizontal spatial resolution, e.g.\ $\Delta x=0.5$~mm. The vertical spatial resolution $\Delta y$ is equal to the distance between the holes in the pepper-pot plate and is in our case $\Delta y=2$~mm. The horizontal and vertical angular resolutions $\Delta x'$ and $\Delta y'$ are determined by the linear size of the ccd pixels $\Delta xp=\Delta yp$, the magnification factor $M$ and the drift length $d$ via $\Delta x'=\Delta xp/Md=\Delta y'=0.85$~mrad.

Before calculating the discretized phase-space density $\rho(x_q,y_p,x'_{jq},y'_{ip})$ from the grey-value array elements $G_{ij}^{pq}$ we have to correct $G_{ij}^{pq}$ for the different hole areas in the pepper-pot plate and for the asymmetrical correction function as mentioned in section~\ref{design}. Denoting the area of the pepper-pot hole with pixel index $p$ as $A_p$ (normalized to the largest hole area) and the value of the correction function at position $(x_q,y_p)$ as $T_{pq}$, the corrected grey-value array $H_{ij}^{pq}$ is given by
\begin{equation}
H_{ij}^{pq}=\frac{G_{ij}^{pq}}{A_pT_{pq}}\;.
\end{equation}
The array $H_{ij}^{pq}$ is directly proportional to the density $\rho$ at the phase-space position $(x_q,y_p,x'_{jq},y'_{ip})$ given by equations~(\ref{eq:detplancor}) and (\ref{eq:detplanang}).

To compare the 4D phase-space distribution $\rho(x,y,x',y')$ with simulations and/or other measurements it is more convenient to calculate its various 2D projections, e.g.\ the spatial distribution $\rho(x,y)$, horizontal and vertical emittance distributions $\rho(x,x')$ and $\rho(y,y')$, and the angular correlation distributions such as $\rho(x',y')$. These 2D distributions are obtained by integrating the 4D distribution $\rho(x,y,x',y')$ over the two phase-space coordinates that are averaged out. As an example, consider the 2D horizontal emittance distribution $\rho(x,x')$ which is obtained from the 4D distribution $\rho(x,y,x',y')$ via
\begin{equation}
\rho(x,x')=\iint\rho(x,y,x',y')\,\ud y\ud y'\;.
\end{equation}
\begin{figure}[tb!]
\centering
\includegraphics[width=14.cm]{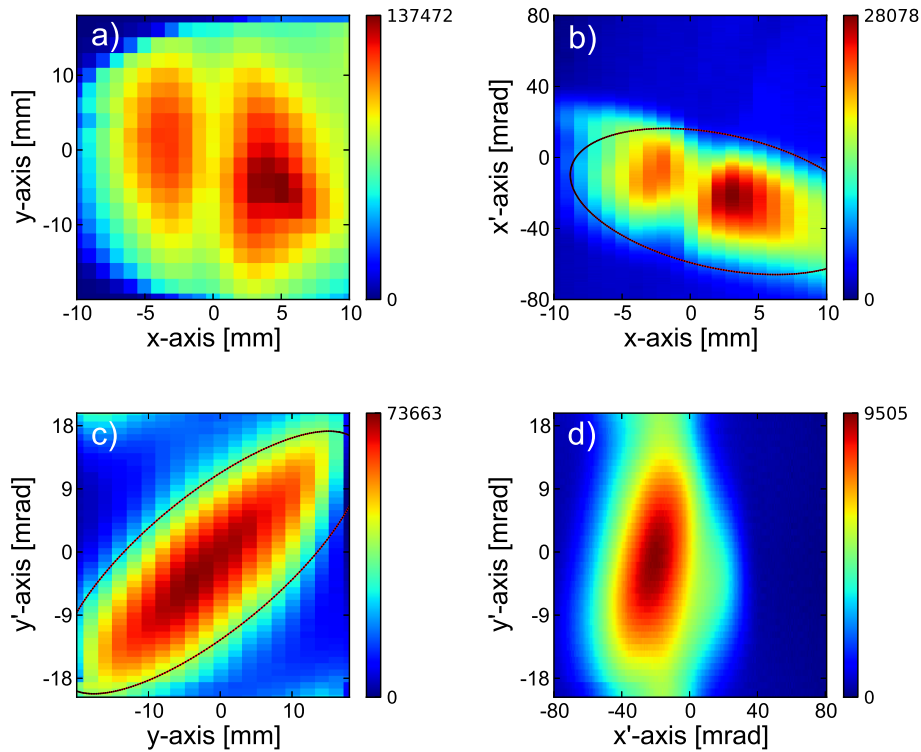}
\caption{\label{fig:fourwin}Two-dimensional projections of the 4D phase-space distribution $\rho(x,y,x',y')$ reconstructed from the data image shown in Fig.~\ref{fig:concat}.}
\end{figure}
The discretized horizontal emittance distribution $\rho(x_q,x'_{jq})$ is obtained from the grey-value array $H_{ij}^{pq}$ by summing the array elements over the indices $p$ and $i$:
\begin{equation}
\rho(x_q,x'_{jq})=C\sum_{ip}H_{ij}^{pq}\;,\label{eq:hjq}
\end{equation}
with $C$ a normalization constant. In equation~(\ref{eq:hjq}) one has to sum over all hole indices $p$ and for each $p$ over all pixel indices $i$ belonging to that value of $p$. The other 2D projections are obtained in a similar way. Fig.~\ref{fig:fourwin} shows four of the six projections calculated from the pepper-pot data file shown in Fig.~\ref{fig:concat}. Note that the $(x',y')$ distribution, which shows a small correlation between the horizontal and vertical planes, cannot be measured with Allison-type emittance scanners.

\subsection{Calculation of beam emittances}
The horizontal and vertical emittance distributions (Fig.~\ref{fig:fourwin}b and c) are conveniently quantified with single numbers, i.e.\ the horizontal and vertical emittances $\epsilon_x$ and $\epsilon_y$, respectively. These are defined using the second-order moments $\sigma_{ij}$ of the corresponding 2D emittance distributions. For example, the second-order moments of the horizontal emittance distribution are given by
\begin{eqnarray}
\sigma_{11} & = & \iint (x-\langle x\rangle)^2\rho(x,x')\,\ud x\ud x' \nonumber \\
\sigma_{22} & = & \iint (x'-\langle x' \rangle)^2\rho(x,x')\,\ud x\ud x' \\
\sigma_{12} & = & \sigma_{21}= \iint (x-\langle x\rangle)(x'-\langle x' \rangle)\rho(x,x')\,\ud x\ud x' \nonumber\;,
\end{eqnarray}
assuming the distribution $\rho(x,x')$ is normalized. The second-order moments $\sigma_{ij}$ define the components of the symmetric horizontal $\mbox{\boldmath{$\sigma$}}_x$ matrix of the ion beam \cite{carey}. The horizontal rms-emittance $\epsilon_x$ is equal to $1/\pi$ times the area of the ellipse determined by the quadratic form $\mathbf{x}^T\cdot\mbox{\boldmath{$\sigma$}}_x^{-1} \cdot\mathbf{x}=1$, with $\mathbf{x}^T=(x,x')$. Thus, the rms-emittance $\epsilon_x$ is also equal to the square root of the determinant of the $\mbox{\boldmath{$\sigma$}}_x$ matrix, i.e.\ $\epsilon_x=\sqrt{\sigma_{11}\sigma_{22}-\sigma_{12}^2}$. Following Lapostolle we quote emittances as four times the rms-emittance, i.e.\ $\epsilon_{x,4rms}\equiv 4\,\epsilon_x$, which encloses 86\% of the particles in Gaussian beams and all particles in uniform beams\cite{lapostolle}.

\begin{figure}[tb!]
\centering
\includegraphics[width=12.cm]{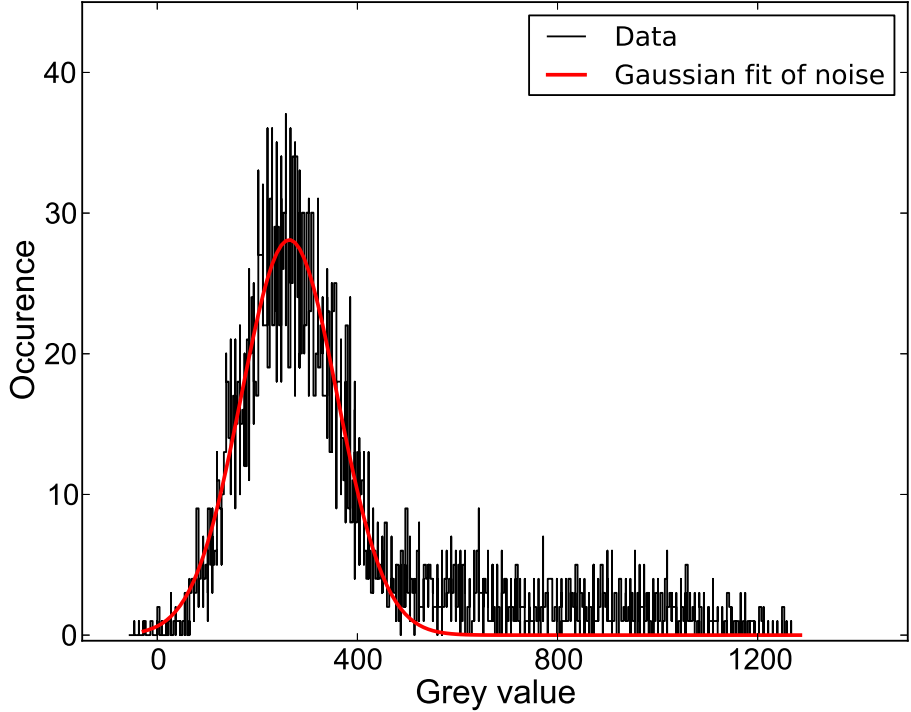}
\caption{\label{fig:gvhistxxp}Gaussian fit through the noise of the gray-value histogram of the horizontal emittance distribution $\rho(x,x')$.}
\end{figure}

\begin{figure}[tb!]
\centering
\includegraphics[width=12.cm]{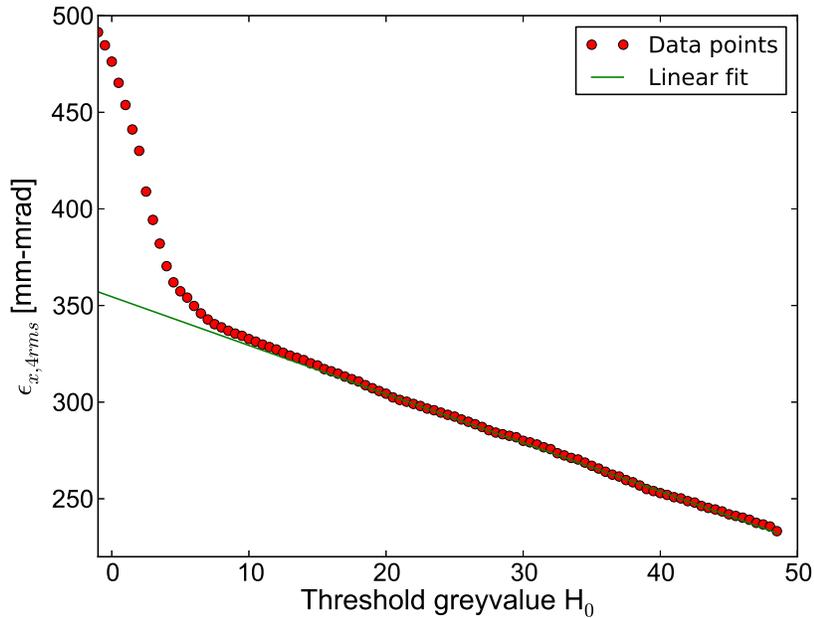}
\caption{\label{fig:em_xxp}Horizontal 4rms emittance $\epsilon_{x,4rms}$ as a function of the grey-value threshold $H_0$.}
\end{figure}

In order to extract reliable values of beam emittances it is important to accurately determine the edge of the beam in the 2D emittance-distribution plots, i.e.\ to separate beam-related grey-value data and noise. The noise is Gaussian distributed as can be seen in the grey-value frequency plot of the horizontal emittance distribution $\rho(x,x')$ shown in Fig.~\ref{fig:gvhistxxp}. The area of the peak is equal to the number of pixels with no beam, the peak width is determined by the ccd noise.
A robust method to separate the noise from the beam-related data points is to define a threshold grey-value $H_0$ and calculate the beam emittance corresponding with all pixels with a grey value $H>H_0$ as a function of $H_0$. This is illustrated for the horizontal emittance distribution $\rho(x,x')$ (Fig.~\ref{fig:fourwin}b) in Fig.~\ref{fig:em_xxp}. As can be seen, the calculated emittance increases sharply for values of $H_0\lessapprox 5$ and decreases linearly for $H_0> 10$. A good estimate of the 4rms-emittance $\epsilon_{x,4rms}$ is obtained by fitting a straight line through the linear portion $(H_1,H_{max}$ of the data points in Fig.~\ref{fig:em_xxp} and extrapolating the linear fit to $H_0=H_{min}$, with $H_{min}$ and $H_{max}$ the minimum and maximum grey values in the data set. The value of $H_1$ is determined such that the standard deviation of $\epsilon_{x,4rms}$ is minimal. This gives a 4rms-emittance of $\epsilon_{x,4rms}=358\pm 8$~mm-mrad for the horizontal emittance distribution of Fig.~\ref{fig:fourwin}b. In the same way we find for the vertical emittance distribution shown in Fig.~\ref{fig:fourwin}c a 4rms-emittance of $\epsilon_{y,4rms}=294\pm 16$~mm-mrad. This procedure only works when the reconstructed emittance distributions contain a sufficient amount of background cq.\ noise. If this is not the case one should repeat the measurement with suitably modified scan and binning parameters and/or change the focussing of the ion beam.


\section{\label{conclusion}Conclusions}
We have designed and constructed a new pepper-pot emittance meter for low-energy heavy-ion beams. Our design differs from existing pepper-pot emittance meters in that we step a pepper-pot plate with a linear array of holes through the beam instead of intercepting the beam with a pepper-pot plate having a 2D array of holes. As 2D ion detector we use a system of MCP's, phosphor scintillation screen and a ccd camera.

An important advantage of our scanning technique is that the spatial resolution in the scanning direction is determined by the step size and/or the spatial resolution of the ion detector and not by the hole pattern in the pepper-pot plate. One should therefore mount the emittance meter at a location in the beam line where the beam width is narrow in one (the scanning) direction and much larger in the transverse direction, e.g.\ near the focus of a quadrupole element. Reconstruction of the transversal phase-space distribution is straightforward as long as beamlets emerging from different holes in the pepper-pot plate do not overlap. When the beam divergence is so large that different beamlets do overlap, phase-space reconstruction is still possible in favourable circumstances but becomes much more elaborate\cite{kremers1}. In these cases it is better to use a  pepper-pot emittance meter which scans the pepper-pot plate in both horizontal and vertical directions.

We have illustrated the operation of our pepper-pot emittance meter using a 15~keV He$^+$ beam at the EIS test bench of GSI, Darmstadt. Extensive comparisons with Allison emittance scanners have been performed at the test bench of A-Phoenix of LPSC Grenoble and can be found in Ref.\ \cite{kremers2}.

\begin{acknowledgments}
This work is part of the research program of the "Stichting voor Fundamenteel Onderzoek der Materie" (FOM) with financial support from the "Nederlandse organisatie voor Wetenschappelijk Onderzoek" (NWO). It is supported by the University of Groningen and by the "Gesellschaft f\"ur Schwerionenforschung GmbH" (GSI), Darmstadt, Germany. Financial support of the European Union Seventh Framework Programme FP7/2007- 2013 under Grant Agreement n° 262010 - ENSAR is gratefully acknowledged. We also would like to thank our colleagues of the KVI design, mechanical and IT departments for their contributions and our GSI colleagues for their help with the measurements.
\end{acknowledgments}

\end{document}